# Scalable, universal and conformal direct electrodes microprinting for high-performance van der Waals-integrated two-dimensional electronics and flexible applications

Nan Cui[1,2,†], Tinghe Yun[1,2,†], Bohan Wei[1,3], Yang Li[1,4], Wenzhi Yu[1,2], Denghui Yan[5], Lianbi Li[5], Haoran Mu[1], Weiqiang Chen[3], Guangyu Zhang[1,2,*] and Shenghuang Lin[1,*]

[1]Songshan Lake Materials Laboratory, Dongguan 523808, Guangdong, P. R. China

[2]Institute of Physics, Chinese Academy of Science, Beijing, 100190, P. R. China.

[3]College of Biophotonics, South China Normal University, Guangzhou 510631, P. R. China.

[4]School of Physics and Materials Science, Guangzhou University, Guangzhou 510006, P. R. China

[5]School of Science, Xi'an Polytechnic University, Xi'an, 710048, P. R. China.

\* Corresponding author email: gyzhang@sslab.org.cn; linshenghuang@sslab.org.cn.

† Theses authors contribute eqaully to this work: Nan Cui, Tinghe Yun.

**Abstract**

Two-dimensional (2D) materials with extraordinary electrical properties, hold promising for large-scale, flexible electronics. However, their device performance could be hindered due to the excessive defects introduced via traditional electrode integration processes. Transfer printing techniques have been developed for van der Waals contacts integration, while existing techniques encounter limitations in achieving conformal electrode transfer and compatibility with flexible devices. Here we introduce a highly conformal microprinting technique utilizing polypropylene carbonate (PPC)/Polyvinyl alcohol (PVA) copolymer, which enables successful transfer of wafer-scale, micropatterned electrodes onto diverse substrates, including those with complex geometries. This technique, implemented with 2D transition metal dichalcogenides (TMDCs), yields 2D field-effect transistors with near-ideal ohmic contacts, and a record-high carrier mobility up to 334 $cm^2$ $V^{-1}$ $s^{-1}$ for a $WSe_2$ device. Furthermore, we fabricated transistor arrays on $MoS_2$ thin film, which show uniform device performance. We also present the flexible $MoS_2$ transistors that not only achieve a high electron mobility of up to 111 $cm^2$ $V^{-1}$ $s^{-1}$ but also exhibit outstanding mechanical robustness. Our findings represent a significant leap forward in the fabrication of flexible 2D electronics, paving the way for numerous emerging technologies.

**Introduction**

Scalable, high-performance electronics that adaptable to soft, non-planar surfaces are crucial for the modern semiconductor industry and various innovative applications, including wearable sensors, flexible displays, skin electronics, and deformable optoelectronics[1-6]. 2D materials, notably TMDCs, are promising candidatures due to their atomically thin feature, dangling bond-free surfaces, and exceptional electrical properties even at atomic-scale thickness[7-9]. Their reduced short-channel effects make them suitable for highly miniaturized transistors, contributing to ongoing device scaling in pursuit of Moore's law[10,11]. Additionally, recent advancements in wafer-scale synthesis techniques enable the fabrication of electronic-grade devices[12,13].

However, the performance of 2D field-effect transistors (FETs) is significantly affected by the metal-semiconductor (M-S) contacts. Conventional electrode integration, utilizing lithography and metal deposition, introduce additional defects, strain, and even damage to the 2D materials[14-16]. This results in a high contact resistance and Fermi-level pinning (FLP) effect at the M-S interface[14,17,18], which greatly limits the overall performance of FETs in terms of on/off ratios and carrier mobilities. Progress has been made to improve the quality of M-S contacts by using 2D metal contacts[19,20], 2D/3D stacks[21], and tunnelling contacts[22,23], resulting in the formation of van der Waals (vdW) contacts with low contact resistance, although most of these strategies are non-scalable due to the use of exfoliated materials for contacts.

Direct metal deposition approaches for scalable vdW integration have been explored, including utilizing advanced deposition techniques[24], deposition on a sacrificial buffer layer[25,26], the laboratory evaporation of Indium/Au contacts[27], and semimetal contacts such as bismuth[28] and antimony[29]. Nevertheless, unsolicited contamination and deterioration of the 2D material surface could still happen during the subsequential lithography process[30]. And these approaches are more applicable for traditional rigid electronics, as the metal deposition process typically necessitates planar substrates for deposition, which may not be optimal for all types of flexible substrates[16], especially those with irregular surfaces.

Transfer printing technique, which typically relies on "pick-up-and-release" process with a polymer stamp, has shown promise in creating high-quality vdW contacts[31-36], and recently, advanced in scalable fabrication. Liu et.al explored scalable electrodes delamination by pre-depositing electrodes on graphene surface[37], and more recently, Yang et.al developed wafer-scale electrode "pick-up-and-release" approach with a quartz/PDMS semirigid stamp and a photolithography mask-aligner[38]. However, these approaches are yet to be explored or implemented for flexible devices. This limitation stems from the polymer stamps, mostly elastomers, possess poor conformality due to their high thickness (> 100 μm)[39], hindering the conformal transfer of microstructures onto irregular and flexible surfaces. While Daus et al. demonstrated a large-scale $MoS_2$ flexible device via transfer with embedded contacts in a polyimide substrate[40], a direct and conformal electrode transfer for flexible 2D electronics, especially for high-performance flexible 2D FETs, remains elusive.

Herein, we design a wafer-scale, conformal and universal direct microprinting strategy for the transferring micropatterned electrodes, facilitated by a co-polymer layer of PPC/ PVA (with the thickness less than 1 μm). Unlike pure PPC, which has an ultralow glass transition temperature ($T_g$), the PPC/PVA boasts an elevated $T_g$ that significantly exceeds room temperature, ensuring robust mechanical stability. Leveraging the thermal fluidity of PPC, the PPC/PVA can transition between solid and viscous liquid states as required, making it an ideal material for high-fidelity, ultra-conformal printing across extensive areas. We have successfully transferred universal micropatterned electrodes (metal, metal nanowires, and transparent conductive polymer) onto various substrates including rigid silicon, flexible substrates, and non-uniform curved surfaces with seamless conformal attachment. By implementing this technique with TMDC materials, we have demonstrated that the $WSe_2$ FETs with Au-vdW contacts exhibit a pronounced p-type behavior and excellent electrical performance, with a contact resistance ($R_c$) of 5 kΩ μm, on/off ratio of $10^7$, and a field-effect hole mobility at room temperature reaching up to 334 $cm^2$ $V^{-1}$ $s^{-1}$. Additionally, we have created $MoS_2$-based FET arrays, showcasing good device-to-device uniformity, the average field-effect electron mobility of 58 $cm^2$ $V^{-1}$ $s^{-1}$, and the on/off ratio of $10^7$. Moreover, we have fabricated the high-performance flexible $MoS_2$ FET with the highest electron mobility up to 111 $cm^2$ $V^{-1}$ $s^{-1}$ and on/off ratio of $10^7$. The device electrical performance is perfectly preserved under strain tests and exhibits a robust tolerance to cycled bending tests.

**Wafer-scale electrodes microprinting**

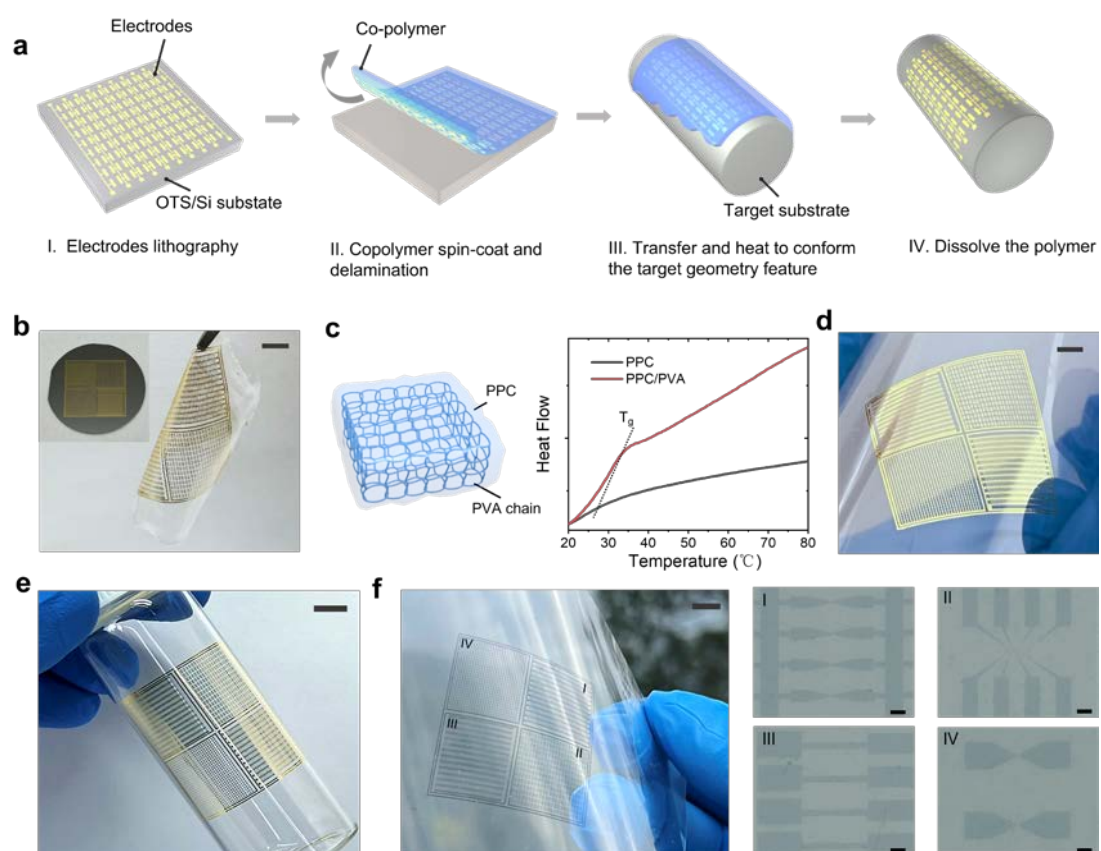

**Fig. 1 Wafer-scale electrode microprinting technique. a,** Schematics of transfer printing steps for non-planar substrate transfer. **b,** Schematic of PPC/PVA cross-linked microstructure (left panel) and the DCS test results for PPC and PPC/PVA, respectively (right panel). **c,** Photographs of a free-standing 2-inch wafer scale PPC/PVA-metal electrode stack. Scale bar, 0.5 cm. **d,e,** Photographs of 2-inch wafer scale metal electrodes transferred onto flexible PET substrate (d) and curved glass bottle surface (e). Scale bar, 0.5 cm. **f,** Photograph of transferred 2-inch wafer scale transparent polymer electrodes on PET substrate with diverse patterns as depicted in region I-IV and the corresponding optical microscopy images. Scale bar (left), 0.5 cm. Scale bar (right), 50 μm.

The microprinting of wafer-scale electrodes involves the electrodes prefabrication (I), delamination with PPC/PVA (II), alignment to the target substrate at elevated temperature (III) and detachment for integration (IV, Fig. 1a). The micropatterned electrodes were fabricated on Si substrate treated with octadecyl-trichlorosilane (OTS) self-assembled monolayers (SAMs)[41], which reduced vdW connection between the electrodes and substrate, ensuring high reproducibility and sample yield. Scalable deposition techniques, including thermal evaporation and solution-processed coating, as well as industrialized lithography were adopted to fabricate the high-precision microelectrodes. More details can be found in Methods.

Notably, the PPC/PVA co-polymer layer plays an essential role in the large-scale conformal electrode printing. While PPC has been used in the localized transfer of 2D materials[42] and 3D metal electrodes[43] due to their low $T_g$ and viscous property, the segment motion of PPC can result in the deformation or cracking of electrodes over a large area (see demonstration in Supplementary Fig. S1). To enhance their mechanical stability, PVA with long cross-linked chains and multi-hydroxyl groups was blended with

PPC, forming a unique network by PVA chains bonding with carbonyl groups of PPC, as schematically illustrated in the left panel of Fig 1b. As a result, $T_g$ is significantly improved from 27 °C for pure PPC to 35 °C for PPC/PVA co-polymer (right panel of Fig. 1b), which is in good agreement with previous studies[44].

With the spin-coating of PPC/PVA on the prepared electrodes, the entire structure (PPC/PVA-electrode stack) can be delaminated after curing the co-polymer at 100 °C for 60s. Subsequently, the free-standing PPC/PVA-electrode stack (Fig. 1c) can be brought in contact with the target substrate on one side. The temperature is then raised to 100°C, well above the $T_g$, which allows the co-polymer to flow and conform the geometric features of the whole substrate. After cooling to room temperature, the resulting solid co-polymer can be dissolved by organic solvents. The corresponding photographs for each step are shown in Supplementary Fig. S2, with the demonstration of transfer wafer-scale Au electrodes with different micropatterns on $SiO_2$ substrate. Supplementary Fig. S3 presents the details of transferred patterns without showing cleavages and cracks of the electrode, illustrates the robustness of this transfer. This microprinting process can be also implemented with a commonly seen house-built transfer set up for deterministic transfer, which requires an additional glass/PDMS stack for mechanical support (see more details in Supplementary S4).

The flexibility and universality of the microprinting approach are well demonstrated with different electrode systems. As a proof of concept, we successfully transferred 2-inch wafer-scale Au electrodes on virous non-planar substrates, including flexible polyethylene terephthalate (PET) substrate (Fig. 1d), and the glass bottle (Fig. 1e). Furthermore, this microprinting is not limited to any specific metal contacts, and theoretically, it is applicable to any choices of conductive materials work as electrodes, including conductive polymers and metal nanowires. To demonstrate this, we successfully execute the transfer of metal electrodes made with Al, Ag and Cu on $SiO_2$ substrates (Supplementary Fig. S5), Ag nanowire (NW) electrodes on artificial leather surface (Supplementary Fig. S6), and transparent polymer (poly(3,4-ethylenedioxythiophene):poly(styrene sulfonate), (PEDOT:PSS)) electrodes on PET substrate (left panel of Fig. 1f). The details of transferred polymer patterns are shown in the optical microscopy images in the right panel of Fig. 1f, which manifest the non-destructive and precise of this transfer.

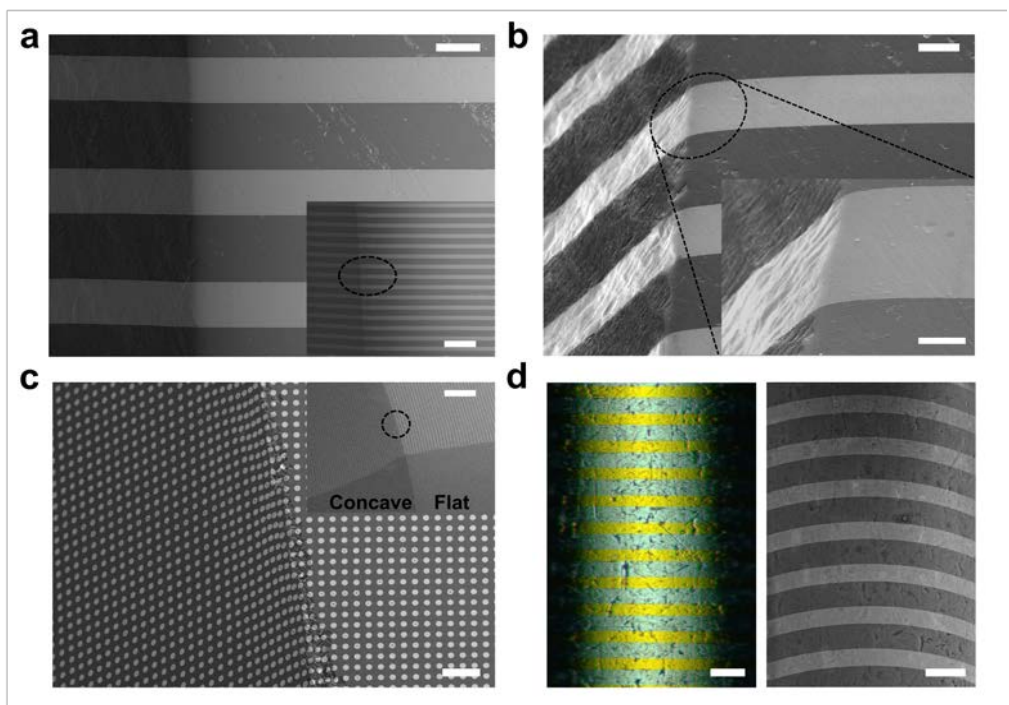

**Fig. 2 Morphology of high-curvature conformal microprinting. a,b,** SEM images of strip electrode arrays transferred onto the cubic corner (a) and tilted with 52 degrees (b). Scale bar (a), 20 μm. Scale bar (b), 10 μm. Inset in (a) provides a lower magnification view of electrode array. Scale bar, 150 μm. Inset in (b) further magnifies the strip over the edge. Scale bar, 5 μm. **c,** SEM image of the disk electrodes transferred onto a concave surface with cracked edges. Scale bar, 30 μm. Inset shows concave structure with a half-concaved and half-flat surface. Scale bar, 100 μm. **d,** Optical and SEM images of strip electrodes transferred onto a syringe needle tip. Scale bar (left), 40 μm. Scale bar (right), 50 μm.

To demonstrate the feasibility of our electrode microprinting for conformal electronics, we transferred arrays of micro-sized strip and disk thin electrodes onto non-planar substrates with non-uniform surface finish. Although thin metal strips are not stretchable, the reflow of co-polymer still guide them bending to conform. As demonstrated in Fig. 2a, the transferred strip array on a cubic corner is clean and continuous around the edge without showing cracks and dislocation, suggesting possibilities of this transfer for practical microfabrication in the semiconductor industry. With a tilt of 52 degrees (Fig. 2b), strips are shown to be attached to the non-uniform surface with excellent conformality since the geometry features of the bottom substrate are evident. Fig. 2c shows an array of disks transferred onto a concave surface (half concave and half flat, as shown in the inset of Fig. 2c) with uniform spacing and neat arrangement, indicating that the reflow of co-polymer is thermally driven without bending and stretching the polymer layer. Despite that the edge of the concave surface is noticeably fractured, it is still uniformly covered by the transferred disks, highlighting this transfer is highly conformal. This microprinting approach also accommodates sharp, tightly curved features, as shown in Fig. 2d, where a strip array is conformally transferred onto a syringe needle tip, implying this technique is applicable to a wide range of objectives from fine textures to macroscopic ones.

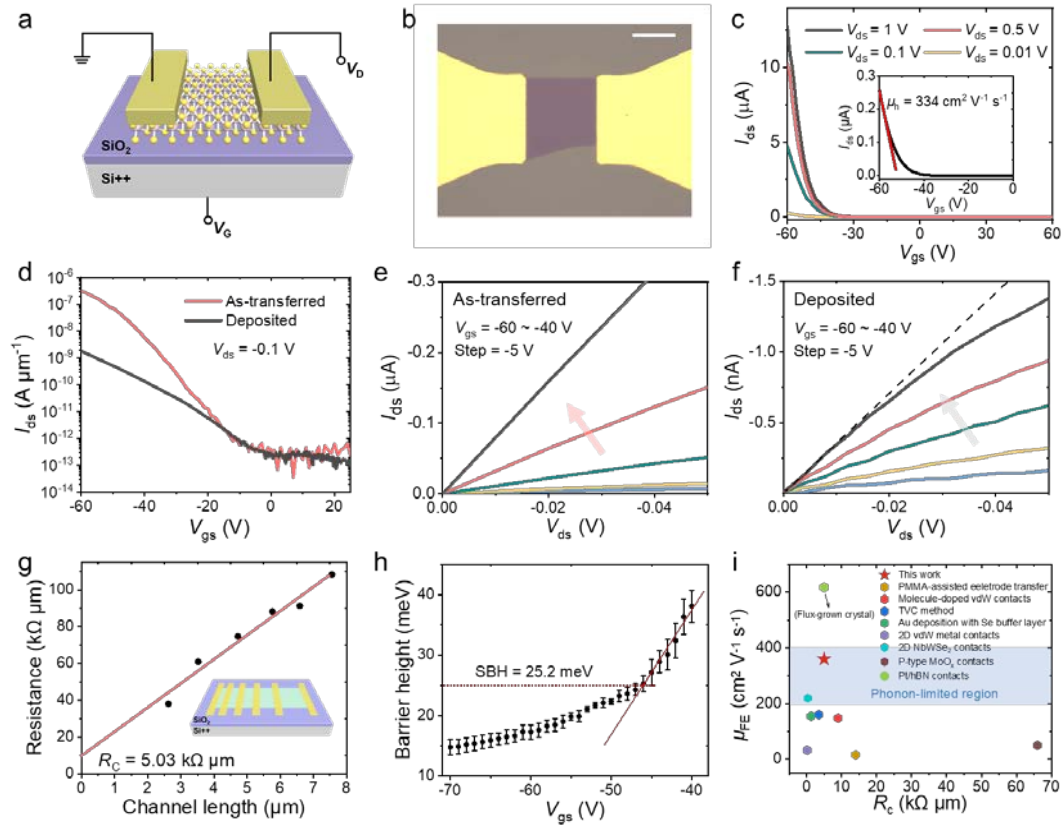

**Fig. 3 Electrical performance of back-gated WSe$_2$ transistors with transferred and deposited metal contacts. a,** Schematic representation of WSe$_2$ transistor structure. **b,** Optical image of a WSe$_2$ transistor with transferred Au contacts. Scale bar, 10 μm. **c,** Transfer characteristics of the WSe$_2$ transistor with transferred contacts under different $V_{ds}$. Inset shows the transfer characteristic at $V_{ds} = 0.1$ V for the subtraction of field-effect mobility. **d,** Transfer characteristics for WSe$_2$ transistor with transferred contact and deposited contact at $V_{ds} = 0.1$V. **e-f,** P-type output characteristics ($V_{gs}$ = -60 to -40V, step = -5 V) of WSe$_2$ transistor with transferred contacts (**e**) and deposited contacts (**f**) (dashed line represents linear function fitting). **g,** Total resistance versus channel length curve for the determination of contact resistance ($R_c$) by TLM method. Inset is the schematic of sample structure with different channel lengths for TLM measurement. **h,** Extracted p-type effective barrier height at various $V_{gs}$ for determining the flat-band Schottky barrier height (SBH). **i,** Comparison of $R_c$ and mobility with previously reported values for WSe$_2$ transistors with vdW contacts.

The performance of individual TMDC-based FETs was characterized at room temperature to demonstrate the advantages of our contact transfer approach. As schematically illustrated in Fig. 3a, the drain/source electrodes were transferred onto the few-layer WSe$_2$ on SiO$_2$/Si substrate. The device fabrication details can be found in Methods section. Au was chosen as contact electrode due to its excellent stability, good ductility, and proper work function. The device was back gated via the highly doped Si, and SiO$_2$ with 300 nm thickness was used as the gate dielectric. To exclude the current spreading effect[45,46] and prevent deterioration of the WSe$_2$ flake caused by the material patterning process[47,48], the width of the WSe$_2$ flake was intentionally constrained to remain narrower than that of the transistor channel, as demonstrated in Fig. 3b. Specifically, the channel width/length was set as 15/15 μm. The transfer characteristics (drain-source current $I_{ds}$, gate-source voltage $V_{gs}$) for as-transferred WSe$_2$ device (Fig. 3c) depicts a pronounced p-type behavior, which is consistent with the majority carrier type expected in few-layer-WSe$_2$ flakes when interfaced with high work function metals such as Au ($W_F$ = 5.1eV), as determined by their band alignment studies[19,35,49]. This indicates the successful achievement of a satisfactory vdW interface using this microprinting approach[50]. From the transfer curve at $V_{ds}$ = 0.01V, where the transistor operates in the linear regime (see output curve in Supplementary Fig. S7), the field-effect hole mobility ($\mu_h$) is determined as the value of 334 cm$^2$ V$^{-1}$ s$^{-1}$ by using the transconductance method. Detailed calculation methods are provided in Supplementary S8.

For comparison, the electrical performance of the deposited WSe$_2$ FET with the identical channel length ((channel width/length = 10/15 μm, see Supplementary Fig. S9a) was also measured, and the contacts were made by lithography and conventional metal deposition technique. The transfer curves for the deposited WSe$_2$ FET (see Supplementary Fig. S9b) also show p-type characteristic of the device. However, its on-current is notably lower, approximately two orders of magnitude, compared to the as-transferred device, and the subthreshold swing (SS) is also smaller for the deposited transistor (Fig. 3d). Furthermore, the output characteristics analysis shows that the transferred transistor (Fig. 3e) exhibits linear behavior at small bias ranges ($V_{ds}$ = 0 to -0.05 V) across various $V_{gs}$ values. Conversely, the WSe$_2$ device with deposited contacts displays Schottky (nonlinear) behavior in its output curves (Fig. 3f). These results suggesting the formation of near ideal ohmic behavior for the transferred contacts, but a significant FLP effect may form at the WSe$_2$/Au interface for the deposited device[17,18]. This could be originated from the excessive defects, strain, and chemical bonding caused by the metal electrode fabrication process[16,32]. In addition, the as-transferred device represents reduced (near-zero) hysteresis during forward and backward scanning of $V_{gs}$ from -80 V to 80 V (Supplementary Fig. S10), indicating fewer defect states at the transferred metal-semiconductor (M-S) interface[51].

To quantify the quality of the as-transferred Au/WSe$_2$ interface, we measured the $R_c$ by transfer length method (TLM) with the configuration shown in the inset of Fig. 3g. The measured $R_c$ for the as-transferred device is determined as 5.03 kΩ μm (Fig. 3h), which is around 9 times smaller than that of deposited contacted device (48.3 kΩ μm, see Supplementary Fig. S11). The Schottky barrier height (SBH) between contacts and 2D material flake was also extracted from Arrhenius plots (Supplementary Fig. S12). The p-type region SBH for the as-transferred device was presented as 25.2 meV, which is comparable to those high quality vdW metal contacts[52]. In contrast, the SBH for the as-deposited device was determined as 196.8 meV (Supplementary Fig. S13), which explains the significant FLP effect at the contact interface and the corresponding poor electrical performance.

Fig 3i benchmarks the performance of as-transferred WSe$_2$ FET against other reported vdW-integrated WSe$_2$ devices[19,20,26,33-35,43,53]. Exceptionally, ref [33] used flux-grown WSe$_2$ crystals with much lower defect density, leading to an ultra-high carrier mobility which exceed the phonon-limited region. Among the work with commercial-grade crystals and CVD-grown flakes, our as-transferred device manifests a comparable contact resistance and a record-high field-effect carrier mobility, to the best of our knowledge. Moreover, our microprinting approach is universally applicable to other TMDC materials for high-performance 2D FETs. As demonstrated in Supplementary Fig. S14, high-performance MoS$_2$ FET was fabricated with a field-effect carrier mobility ~ 83 cm$^2$ V$^{-1}$ s$^{-1}$, which is at a similar level with other best-reported vdW-integrated MoS$_2$ devices[54]. Few-layer MoS$_2$ were also contacted with transferred Al, Cu, and Ag metals, as shown in Supplementary S15. The electrical performance of the corresponding FET highlights the reliability of this technique for 2D devices (Supplementary Fig S16).

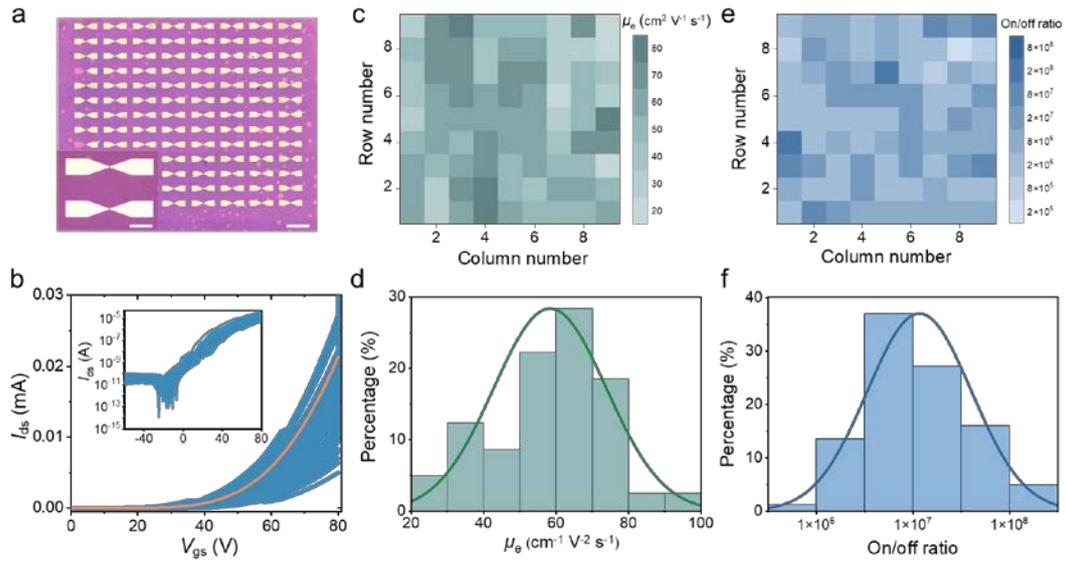

**Fig. 4 Electrical properties of MoS$_2$ FET array with transferred Au contact. a,** optical image of the MoS$_2$ back-gated FET array on SiO$_2$ substrate. Scale bar, 1 mm. Inset is the optical image of two MoS$_2$ FETs. Scale bar, 200 μm. **b,** Transfer curves of 9×9 MoS$_2$ FETs at $V_{ds}$ = 1 V. Inset shows the transfer curves with logarithm y-axis. **c,d,** Mapping (c) and histogram (d) of carrier mobilities of 9×9 MoS$_2$ FETs extracted from transfer curves. **e,f,** Mapping (e) and histogram (f) of on/off ratios of 9×9 MoS$_2$ FETs.

In industrial production, batch fabrication with high consistency and homogeneity is essential. To this end, we prepared FET array based on polycrystalline MoS$_2$ thin film (monolayer, domain sizes around several microns[55]) by batch processing with our transfer printing technique (see fabrication and synthesis details in Method). Fig. 4a depicts the optical image of more than 81 back-gated, Au-contacted FETs with identical channel geometry ($L/W$ = 15/20 μm) fabricated on SiO$_2$ (300 nm)/Si (p++) substrate.

To evaluate the device performance, we measured 9×9 MoS$_2$ transistors in the batch, and the corresponding transfer curves (Fig. 4b) show clear n-type characteristics with small device-to-device variations and high on/off ratios (> $10^6$, inset of Fig. 4b). However, the $I_{on}$ is lower than that of MoS$_2$ FET with exfoliated monocrystalline flakes, which is caused by the significant scattering effect at domain boundaries of the polycrystalline material[56,57]. For all measured devices, the electrical performances are statistically analyzed. Fig. 4c,e demonstrates the uniform distribution of field-effect carrier mobility and on/off ratio with the maximum values reach to 94 cm$^2$ V$^{-1}$ s$^{-1}$ and 4×10$^8$, respectively. The averages are determined from the histogram (Fig. 4d,f with Gaussian fit to the distribution) as 58 cm$^2$ V$^{-1}$ s$^{-1}$ and 2×10$^7$, respectively. These values suggesting the great electrical performance of our devices, which is comparable to the best-reported transistor array based on high-quality CVD- MoS$_2$ with large grain sizes (see comparison in Supplementary S17 and Table S1), emphasizing the superior resultant device quality with our fabrication approach. Therefore, our transfer printing process is shown to be well-suited for batch fabrication of high-performance 2D electronics.

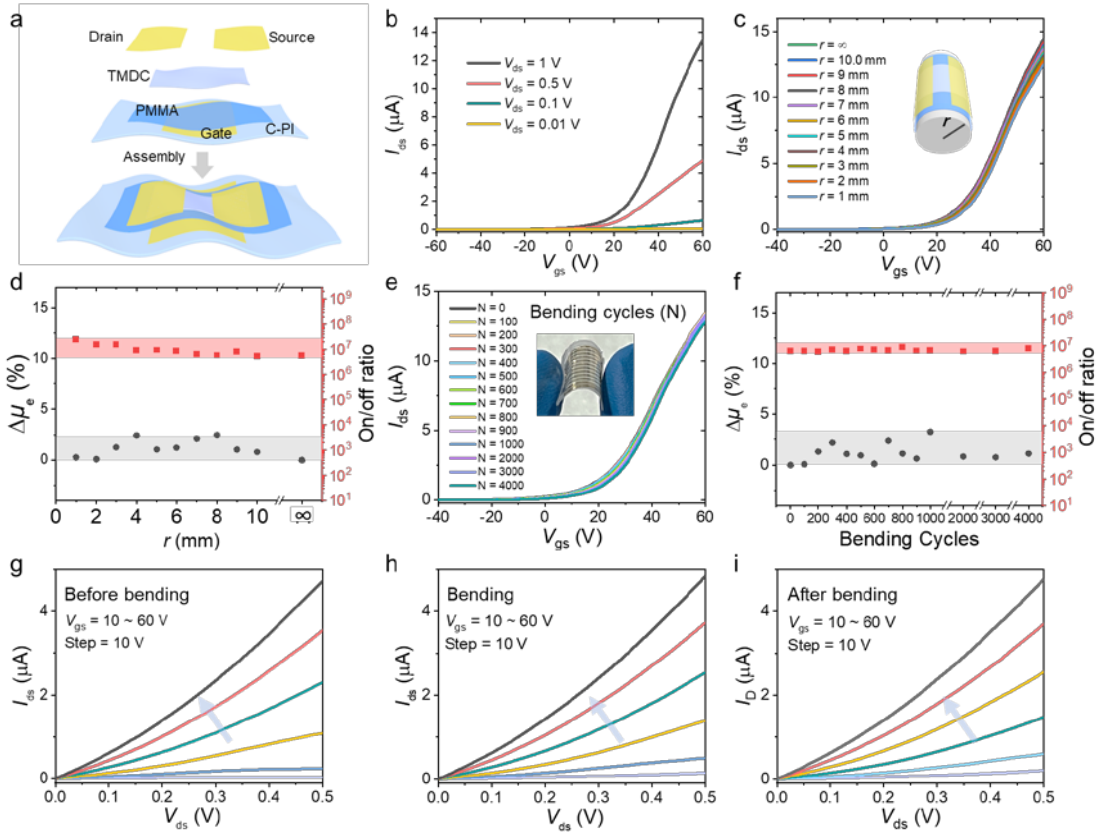

**Fig. 5 Electrical characterization of a flexible vdW-integrated MoS$_2$ device under strain and cycle tests. a,** Schematic illustrating the integration of flexible MoS$_2$ transistors by using the microprinting technique. **b,c,** Transfer characteristics of a flexible MoS$_2$ transistor under different $V_{gs}$ (b) and under varying radii of curvature ($r$) at $V_{ds}$ = 1 V (c). **d,** Mobility deviation and on/off ratio dependence on different bending radii. $\Delta\mu_e = (\mu_e - \mu_{e0})/ \mu_{e0}$, where $\mu_e$ and $\mu_{e0}$ stand for the electron mobility under bending strain and initial mobility, respectively. **e,** Transfer characteristics ($V_{ds}$ = 1 V) of a flexible MoS$_2$ transistor subjected to bending and recovery cycles. Inset shows the photograph of a bending device. **f,** Dependence of mobility deviation and on/off ratio on different bending cycles. **g-i,** Output characteristics ($V_{gs}$ = 10 to 60 V) of a flexible MoS$_2$ device: (**f**) before bending, (**g**) during bending, and (**h**) after bending.

To investigate the microprinting technique for flexible applications, we fabricated MoS$_2$ transistors on flexible substrates, as depicted in Fig. 5a. The gate electrode was transferred onto a flexible colorless polyimide (C-PI) substrate, with PMMA serving as the gate dielectric. The capacitance measurement, detailed in Supplementary Figure S18, was utilized for the mobility calculation. Followed by the transfer of exfoliated MoS$_2$ flakes used as active layer, and drain/source electrodes were then released using the microprinting technique (see structure cross-section in Supplementary Fig. S19, and fabrication details in Methods section). Fig. 5b shows the typical transfer curves for the flexible MoS$_2$ FET with obvious n-type transport behavior under different $V_{ds}$. The statistical electrical performance of flexible MoS$_2$ shows the average electron mobility of 94.2 cm$^2$ V$^{-1}$ s$^{-1}$ with the standard deviation of 13.0 cm$^2$ V$^{-1}$ s$^{-1}$, and highest value up to 111 cm$^2$ V$^{-1}$ s$^{-1}$ (Supplementary S20 and Fig. S21). The above result is superior to other high-performance flexible FETs based on thin MoS$_2$ (see benchmark flexible MoS$_2$ device in Supplementary S22). The transfer characteristics shown in Fig. 5c reveal that the electrical performance remains consistent across a wide range of bending radii $r$ ($r = 1$ to 10 mm), and the field-effect mobility as a function of bending radius, as depicted in Fig. 5d, exhibits a stable trend with negligible deviation of less than 3%. The overlapped Raman spectra in Supplementary Fig. S23 for MoS$_2$ channel before and after bending indicate that the charge transport mechanism within the TMDC channel is largely unaffected by moderate mechanical strain of ⩽0.6%.

Furthermore, to test the longevity and durability of the flexible device, we performed cyclic bending tests (inset of Fig. 5e) up to 4000 times. Fig. 5e shows the transfer curves after various bending cycles remain unchanged, and the steady electron mobility and minimal deviation percentage (Fig. 5f) highlight the device ability to withstand repeated mechanical stress without significant degradation in performance. The output characteristics before, during, and after bending, illustrated in Fig. 5g-i, show an outstanding consistent behavior with the full-recovery of electrical performance, further attesting to the robustness of the flexible device. The van der Waals (vdW) sliding contact may also facilitate the release of the extrinsic strain through weakly coupled metal-2D interfaces[58]. These results analysis confirm the mechanical flexibility and electrical reliability of our flexible device under practical dynamic conditions, suggests the high feasibility and reliability of our microprinting technique for the integration of high-performance flexible electronics.

**Conclusion**

In summary, we introduced a highly scalable, universal, and conformal direct electrodes microprinting technique, enabled by the PPC/PVA co-polymer. By employing this approach with few-layer TMDC flakes, we achieved a near ideal ohmic M-S interface with low contact resistance. The individual FET device on rigid SiO$_2$ substrates exhibits large field-effect carrier mobility and high on/off ratio, and the FET array in large-scale demonstrates uniform device performance. Moreover, we fabricated flexible devices that exhibit great electrical performance, superior resilience under strain, and long-term stability during cyclic testing. Our work provides a robust approach for fabricating high-performance, large-scale, flexible 2D electronics across diverse substrates, which holds significant potential for next-generation flexible electronic applications.

**Methods**

**Polymer preparation**

PPC and PVA were purchased from Sigma-Aldrich. PPC/PVA copolymer was obtained by adding PVA aqueous solution (5 wt%) into PPC solution (20 wt%, anisole) with the volume ratio of 1:1. The polymer solution was stirred on a hot plate at 50 °C overnight to mix the solution thoroughly, then stored in the glove box to exclude bubbles in the mixture. PPC/PVA hybrid solution was spin-coated on the prefabricated microelectrodes at 3000 rpm for 40 s, cured at 100 °C for 30 min.

**Prefabrication of patterned electrodes on OTS/Si substrates**

Si substrate surface was pre-treated by immersing Si substrate into the OTS solution (OTS/n-heptane = 1:2000 by volume) for 20 min. The positive photoresist (AZ 5214 E, Clariant) OTS/Si substrate was patterned via photolithography and developed in AZ 300 MIF. Subsequently, the metal (gold, silver, copper, aluminum) films with the thickness of 50 nm were thermally evaporated onto the photoresist pattern, followed by a lift-off process. As a result, the micropatterned metal electrodes can be obtained on the OTS-treated Si substrate.

For the micropatterned polymer electrode, the PEDOT:PSS aqueous dispersion (Clevios PH1000, Heraeus) with 6 vol% ethylene glycol and 0.1 vol% surface active agent (Capstone FS-30) was spin-coated on the OTS-treated Si substrate[41]. A 15 nm thick gold film was thermally evaporated onto the

PEDOT:PSS film as the protective layer, followed by the patterning of photoresist on PEDOT film with photolithography process. The uncovered gold area was removed by gold etchant (I⁻/I₃, Sigma Aldrich), then the exposed PEDOT was etched by oxygen plasma (100 W, 3 min). Finally, the photoresist and the gold film on the patterned PEDOT was removed by acetone and gold etchant, respectively.

For Ag NW microelectrodes, the Ag NW dispersion (0.25 wt%) was spin-coated on the OTS-treated Si substrate, then patterned photoresist on the Ag NW film by photolithography. The uncovered Ag NW was etched by gold etchant solution, and the unexposed photoresist was removed by acetone, 2-propanol, and deionized water in sequence. The sample was finally dried at 100 °C for 30 min.

**Fabrication of TMDC-based devices**

For the individual TMDC FET, TMDC flakes were exfoliated with a conventional exfoliation process from commercial-grade crystals (sourced from HQ graphene[59]) on PDMS layers, then the exfoliated flakes were transferred onto a clean 300 nm-SiO₂/Si substrate with a typical dry transfer method[60]. Drain/source metal electrodes were then implemented by the deterministic microprinting process with the assistance of co-polymer PPC/PVA.

For the wafer-scale MoS₂ FET array fabrication, polycrystalline MoS₂ thin film was synthesized on 300 nm-SiO₂/Si substrate by CVD method. In specific, Sulfur (S) (99.95%, Aladdin) and MoO₃ (99.95%, Aladdin) were used as precursors and heated to 150 °C and 560 °C for 30 min, respectively. The pressure in the growth chamber was ~1 Torr, and Ar (100 sccm) and Ar/O₂ (80/1 sccm) were introduced for the material synthesis (see more details in ref [55]). Subsequently, large-area drain/source electrode array was transferred on the MoS₂ film with the microprinting technique, and the co-polymer was removed with acetone and 2-propanol bath.

For the flexible MoS₂ FET fabrication, gate electrodes were fabricated on OTS-Si substrate by photolithography. Next, PPC/PVA copolymer was spin-coated onto the prepared gate electrodes and cured at 100 °C. The PPC/PVA-gate electrode was delaminated from the OTS-Si. Subsequently, the free-standing PPC/PVA-electrode stack was brought in contact with the target flexible substrate (commercial C-PI with the thickness of 12.5 µm). The temperature was raised above the $T_g$ of copolymer to make the co-polymer conform with the flexible substrate. After cooling to room temperature, the resulting solid co-polymer can be dissolved by organic solvents (such as acetone, chloroform). Poly (methyl methacrylate) (PMMA) (4% wt in anisole) was spin-coated at 3500 rpm for 60 s onto the C-PI with the gate electrodes, followed by a dry transfer of the MoS₂ flake. The drain/source electrodes were ultimately released onto the MoS₂ flake using the microprinting technique. The entire stack was placed onto a curved substrate and tested the electrical performance of the FETs under various bending strains. The bending strain could be calculated from the equation $\varepsilon = \tau/2r$, where $\tau$ and r are the substrate thickness and curvature radius, respectively.

**Material characterization and device measurement**

The DSC test was performed by a differential scanning calorimeter (NETZSCH DSC 200F3) under a nitrogen atmosphere with the heating/cooling rate of 10/20 °C min⁻¹. The optical microscopy images were obtained using an Olympus microscope (BX53M) and a confocal microscope (OLS5000-SAF). SEM images were obtained from high-resolution field-emission scanning electron microscope (ThermoFisher Verios 5UC). The electrical characteristics of the 2D FETs were recorded with Keysight B1500A semiconductor device analyzer in a vacuum probe station under the pressure of 10⁻³ Torr. The electrical characteristics of flexible devices under various bending strains were tested at room temperature under ambient conditions.

**Data availability**

All data supporting the findings of this study are available from the corresponding author upon reasonable request.

**Acknowledgement**

We acknowledge the support from the National Natural Science Foundation of China (No. 62204165), Guangdong Basic and Applied Basic Research Foundation (No. 2021B1515120034), National Key R&D Program of China (No. 2021YFA1202902), and National Natural Science Foundation of China (No. 12204336).

**Author contributions**

S.L., N.C., and T.Y. conceived the ideas and designed the project. G.Z. and S.L. directed and supervised the research. N.C. and T.Y. introduced the PPC/PVA copolymer for conformal electrode transfer, and fabricated and characterized the devices. N. C., T.Y., B.W., and Y.L. conducted electrode and 2D materials transfer. N.C., T.Y., W.Y., and H.M. performed the DSC, SEM, and electrical measurements. D.Y. and L.L. synthesized $MoS_2$ thin film by CVD method. N.C., T.Y., and B.W. fabricated the flexible devices and performed the electrical characterization during mechanical bending tests. N.C., T.Y. and S.L. co-wrote the manuscript. All the authors discussed and analyzed the results in the manuscript.

**Competing interests**

The authors declare no competing interests.